\documentclass[10pt,journal,compsoc]{IEEEtran}

\usepackage{graphicx}
\usepackage{xcolor}
\usepackage{xspace}
\usepackage{comment}
\usepackage{framed}
\usepackage[strict]{changepage}

\usepackage{todonotes}

\usepackage{booktabs}
\usepackage{multirow}

\newcommand{\rqone}{How do developers with disabilities perceive their inclusion in the training camps?\xspace}
\newcommand{\rqtwo}{What challenges do developers with disabilities face when participating in training camps?\xspace}

\definecolor{formalshade}{rgb}{0.95,0.95,1}

\newenvironment{formal}{%
  \MakeFramed{\advance\hsize-\width\FrameRestore}%
  \noindent\hspace{-4.55pt}
  \begin{adjustwidth}{}{7pt}%
  \vspace{2pt}\vspace{2pt}%
}
{%
  \vspace{2pt}\end{adjustwidth}\endMakeFramed%
}

\newcommand{\totalInverviews}{12\xspace}

\begin{document}


\title{Supporting the Careers of Developers with Disabilities: Lessons from Zup Innovation}

\author{Isadora Cardoso-Pereira, Geraldo Gomes*, Danilo Monteiro Ribeiro*,  \\
Alberto de Souza, Danilo Lucena, Gustavo Pinto \\ 
Zup Innovation, Brazil *SENAC - PE, Brazil

}


\IEEEtitleabstractindextext{%
\begin{abstract}
People with still face discrimination, which creates significant obstacles to accessing higher education, ultimately hindering their access to high-skilled occupations. 
In this study we present Catalisa, an eight-month training camp (developed by Zup Innovation) that hires and trains people with disabilities as software developers. 
We interviewed 12 Catalisa participants to better understand their challenges and limitations regarding inclusion and accessibility. 
We offer four recommendations to improve inclusion and accessibility in Catalisa-like programs, that we hope could motive others to build a more inclusive and equitable workplace that benefits everyone.


\end{abstract}

}

\maketitle

%
\IEEEpeerreviewmaketitle

\section{Introduction}
%
Despite many countries implementing regulations to reserve jobs for people with disabilities, they still face discrimination, which creates significant challenges in accessing job opportunities. 
These challenges stem from various factors, including educational barriers that make it less likely for them to pursue higher education, resulting in only around one-third of working-age individuals with disabilities employed globally (half the proportion of people without disabilities)\footnote{https://ilostat.ilo.org/new-ilo-database-highlights-labour-market-challenges-of-persons-with-disabilities/}. 
Moreover, this educational gap reduces their chances of working in high-skilled occupations, such as software developers. 
In the 2022 StackOverflow survey (N=70,000), only $3.6\%$ identified themselves as people with disabilities\footnote{https://survey.stackoverflow.co/2022/\#developer-profile-demographics}.

One way to minimize this gap is through \textit{training camps}, which offer 
an intense and structured process that involves classes, practical training, project work, and mentorship over a few weeks or months, and work as a gateway to the job market by teaching individuals with little to no experience.
However, training software developers with disabilities can pose significant challenges due to the need for innovative teaching methodologies that address accessibility concerns~\cite{moster2022can}. 
For instance, visually impaired developers may face learning barriers because of the visual cues that educational materials and programming tools often rely on~\cite{huff2021exploring,mealin2012exploratory}, while neurodivergent individuals may struggle with communication and teamwork skills~\cite{moster2022can,begel2021remote}. 
By overcoming these challenges, we can harness the benefits of diversity, including the ability to solve corporate and software problems more effectively, resulting in technological advancements and increased competitiveness~\cite{Vasilescu2015Perceptions}. Furthermore, a diverse workforce provides a more enriching, understanding, and learning experience for everyone involved~\cite{Bosu2019Diversity}.



In this work, we present a set of lessons learned through Catalisa, a program aimed to improve the plurality of people and ideas at Zup Innovation.
Catalisa is a eight-month training camp, with immersive technical and soft skills training for people with disabilities. 
By interviewing \totalInverviews Catalisa participants, we shed some light on the challenges and limitations that the participants faced regarding inclusion and accessibility. 
Based on our findings, we propose four recommendations to further improve inclusion and accessibility in workplaces. 
We hope our work can help build a better workplace with more diversity.

\section{Catalisa}\label{sec:catalisa}


Catalisa\footnote{https://www.zup.com.br/zup-academy/catalisa} is a training program 
launched by Zup Innovation, a Brazilian tech company striving to transform Brazil into a global technology hub with diverse teams. 
Launched in 2021, the Catalisa team has already organized and delivered four successful training camps, with 55 participants.  
The program's effectiveness is reflected in the fact that $85\%$ of the participants who completed the program still work at Zup; the program has increased the number of people with disabilities at the company by $45\%$. 
%
Zup invests in one instructor, a manager, a Tech Lead, and a Libras (Brazilian Sign language) interpreter to run Catalisa. 
They share this structure with other training camps for social minorities, such as low-income individuals. 

Catalisa selects candidates through a four-step process: 
registration, online test, short course, and interviews. 
Those who completed these stages attended an onboarding process, where they were introduced to Zup's operations. 
To ensure participants could entirely focus on learning, Zup hired them as trainee software developers from the outset of the training, which also guaranteed their labor rights. 
Additionally, Zup provided the necessary materials to ensure their success, such as personal computers and adaptations based on each person's needs (e.g., large monitors and screen readers).

\subsection{Training process}

Catalisa had an eight-month training phase where one instructor teaches topics related to Java development, starting with programming logic and advancing to Restful API development, unit tests, and database interaction. 
Also, each participant had individual mentoring to support their learning. 
Mentors and Zup's Business Partners also provided workshops and individual mentoring about behavioral skills such as communication, feedback, and emotional intelligence.
Participants had weekly coding challenges where they could achieve one of three rating levels: minimum, medium, or maximum. 
To continue in Catalisa, they must consistently achieve passing scores, demonstrating their commitment to the program's requirements.

In the final week, participants were separated into teams to develop a functional project with business value. 
The mentors chose team members based on their strengths and weaknesses, creating balanced teams in technical and behavioral aspects. 
Mentors and instructors acted as product owners, guiding participants to use agile methodology practices like dailies and retrospectives. 
The final project was a full-stack project written in HTML, CSS, and Javascript on the front-end, and Java with Spring on the backend, also with database integration and unit tests.  
The participants presented their final project as a pitch to Zup's tech leads, showcasing the skills they learned during Catalisa through these projects. 

\section{Method}


\subsection{Research Questions}\label{sec:rqs}

We sought to answer the following research questions:

\begin{itemize}
    \item[\textbf{RQ1:}] \rqone
    \item[\textbf{RQ2:}] \rqtwo
\end{itemize}


For RQ1, we aim to gain insights into the participants' experiences regarding their inclusion in training camps. 
Learning a new profession in a condensed timeframe is demanding, and this can be exacerbated for individuals with disabilities if training programs do not consider their accessibility needs. 
Hence, we posed questions that helped us understand how the inclusion measures implemented in the training improved their motivation and engagement.

For RQ2, our goal is to understand the challenges faced by people with disabilities during the training phase. 
We seek to identify in our interviews any barriers that may have hindered their learning experience and gather feedback on further improving the inclusion measures.



\subsection{Interviews}\label{sec:interviews}


We conducted \emph{semi-structured} interviews, in which the interviewer had a set of questions but could freely modify the questions as needed. 
We interviewed a total of \totalInverviews Catalisa participants to identify practices that other companies could adopt to create more inclusive training programs. 
We also aimed to determine areas for improvement in Catalisa's inclusivity efforts. 
We conducted the interviews and the subsequent analysis in Portuguese; we translated the quotes in this paper.

\subsubsection{Interview questions}


The authors designed
a single script that best answered our RQs. 
We subsequently refined the script based on feedback from two pilot interviews, identifying areas for improvement, such as better wording for certain questions. 
We did not include these pilot interviews in our analysis. 
The resulting interview script comprised 25 questions, including:

\begin{itemize}
     \item Tell us about your work day/study during Catalisa?
     \item Did you experience difficulties in any teaching practice?
     \item Could you describe a situation in which you felt satisfied during the training? And frustrated?
\end{itemize}


\subsubsection{Participants recruitment and interview procedure}

We utilized a \textit{snowball sampling} strategy to recruit our study participants, beginning with one participant (selected based on proximity) and then asking each the interviewee to suggest additional individuals until we could identify no further participants. 
Out of the 14 participants we invited, 12 participated in the interviews. 

While we knew all interviewees had at least one disability (as they had participated in Catalisa), we did not inquire about their specific disability during the interview to avoid being insensitive or inappropriate, focusing on their accessibility needs to work autonomously. 
However, they could disclose relevant information, including their disability status, during interviews, knowing we will respect their privacy. 
Only three participants chose not to disclose, while nine did, including four with visual impairments (total blindness or low vision), three with physical impairments, and one neurodivergent (autism or ADHD).

Before conducting interviews, we informed participants that we would record interviews with their consent, ensuring the anonymization and confidentiality of data,  limitation of data usage to this study only, and the right to refuse to answer questions or stop the interview at any time (in fact, two participants exercised this right and paused the interview, resuming it the next day). 
We also explained the study's goals and potential social benefits.

We recorded all interviews 
and later translated them for data analysis. 
We also took notes on the participants' behavior during the interviews to aid our coding process. 
Four authors conducted the interviews, with at least two authors present for each interview. Throughout this paper, we refer to the participants as P1-P\totalInverviews. 
The interviews, on average, lasted 57 minutes (min: 33, max: 103).

\subsubsection{Data transcription and analysis}


After the interviews, we transcribed the audio using Amazon Transcribe\footnote{https://aws.amazon.com/pt/transcribe/}, 
an automatic service to translate speech to text. 
One author refined the transcripts manually.




\begin{figure}
    \centering
    \includegraphics[width=1\linewidth]{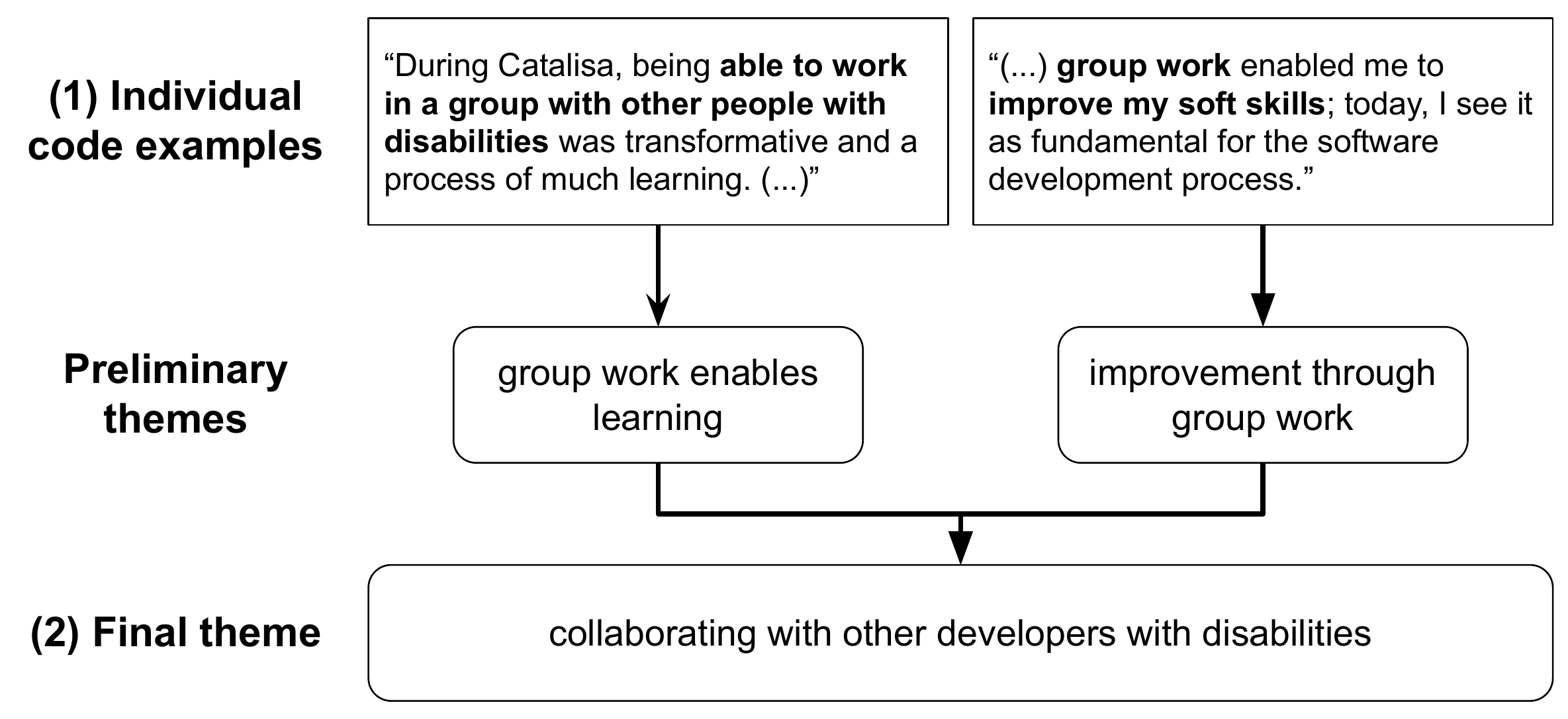}
    \caption{Example of code in transcription for category \textbf{collaborating with other developers with disabilities}}
    \label{fig:code_transcript}
\end{figure}

Our coding process is illustrated in Figure~\ref{fig:code_transcript}. 
\textbf{(1)} Three authors independently performed open coding on the transcripts using general guidelines~\cite{seaman1999qualitative}. 
This process involved creating \textit{preliminary themes} for significant text segments that addressed our RQs. 
\textbf{(2)} We then compared our findings, discussing any discrepancies until reaching agreement on the categorizing themes into broader and \textit{final themes} that represented the perceptions of Catalisa's participants regarding inclusion. 
This multi-perspective approach helps minimize bias and increase confidence in our results.

\subsection{Limitations}\label{sec:limitations}


Our main limitation is the small sample size, which may affect the generalizability of our findings to other companies. 
It is also essential to note that our study focused only on training camps and remote work, which presents unique challenges and may not reflect the experiences of developers with disabilities in other contexts, such as on-site work.

Furthermore, disability is a broad spectrum encompassing many conditions, and individuals with the same impairment may have distinct experiences and needs. However, our research provides valuable insights into the experiences and needs of developers with disabilities, which can inform further actions to promote inclusion and accessibility in workplaces.



\section{Results}

\subsection*{RQ1: \rqone}

The participants emphasized that \textbf{accommodating their unique learning needs} creates a more welcoming and inclusive training environment. 
For example, some participants preferred the freedom of turning the camera and microphone on or off during classes, which is a critical aspect of the remote learning process~\cite{alim2022does,schwenck2021student,castelli2021students}. Our work specifically aimed to examine the accessibility considerations associated with this choice. 
P7 felt that this made the learning experience more inviting and inclusive, e.g., \textit{``we didn't need to open the camera, we didn't necessarily need to talk, we could send questions through the chat, so it was very welcoming.''}. 
Similarly, P1 felt that the mentors created a comfortable learning environment in which participants could make their own choices: \textit{``participating in classes and projects daily was quite natural.''}. 

In contrast to some previous studies~\cite{alim2022does,schwenck2021student} where turning the camera on was found to increase focus, P2, for instance, felt that their interruption could disrupt the learning experience, using the example of installing Git during class. \textit{``If you install it live with people, they will start to have problems live, they will start to stop the class, interrupt it, and although the classes were recorded, it was impossible to rewatch because the interruption broke the linearity.''}. 
While P1 did not find microphone interruptions a problem, they acknowledged the importance of preserving linearity in learning for visually impaired individuals since they need to memorize ideas instead of relying on visual cues. 
Likewise, P6 found the interruptions 
distracting and challenging to their focus, which can be particularly difficult for a neurodivergent individual~\cite{bi2022accessibility}. 
%


Participants also noted that \textbf{continuously improving the training program's accessibility} is crucial. 
They shared that some instructors lacked experience creating accessible learning materials, which aligns with recent literature~\cite{bi2022accessibility} showing that accessibility is not a well-developed skill in software developers. 
However, the instructors were always receptive to feedback, and the suggestions positively impacted the quality of the lectures. As stated by P11: \textit{``We recommended more detailed descriptions of the images in the presentations, slower explanations, and avoiding phrases like 'on the screen, I'm doing this' and 'click with the mouse here.' Sharing these details with the instructor made the classes better throughout the course. At the end of the training, the professor even thanked us for our feedback, acknowledging that it helped him become a better professional with better classes''}. 
As remarked by P1, P3, and P11, other bootcamps did not consider their needs regarding accessibility, making them feel unwelcome and unable to participate. 
P12 also expressed satisfaction with the open communication channels 
and the willingness of the mentors to address their concerns, allowing for the identification of areas for improvement:  
\textit{``We always had a chance to talk to our coordinator, who would relay our comments to the instructors. 
(...) We observed a continuous improvement in the training, including more standardized presentations''}. 
P1 summarized the importance of continuous communication 
as \textit{``We want Catalisa to achieve great results. We want the next training camp, or even the next class, to be better''}.






Moreover, some participants highlighted the significant benefits of \textbf{collaborating with other developers with disabilities} in their personal and professional lives. 
As P9 noted, \textit{``Being able to work in a group with other people with disabilities was transformative. Our group activities taught us to deal with differences in thinking, behavior, knowledge, always trying to understand each other's needs. For example, 
we would check if the sound was clear before starting a meeting, etc. If there were an image, we would describe it.''}.  
P3 added how it also impacted their personal lives, \textit{``
Now I always think: is this accessible to everyone? The contact with different disabilities completely changed my behavioral view at work and in my personal life.''}. 
Therefore, participants noted the importance of group work, especially for including people with visual impairment. 
P11, who is blind, shared that \textit{``Before Catalisa, it was difficult for me to work in a group. I always felt excluded. My name was often added to group work without contributing because others didn't want to deal with my disability. I built a perception that doing everything alone was better. During Catalisa, group work enabled me to improve my soft skills, and today, I see it as fundamental for my career.''}.

Research has shown that \textbf{recording lectures} can positively impact students' learning by allowing them to study at their own pace~\cite{mackay2019show,nightingale2019developing}. 
Similarly, participants in our study also recognized the inclusiveness benefits of recording lectures for people with disabilities.  
P2 mentioned that \textit{``live class is not always positive since I could not practice''}. 
It occurred because the screen reader needed to read the code to them and, as P1 declared, \textit{``either I listen to the person, or I listen to the screen reader... How am I going to be able to write codes if I'm only going to be able to study the theory, but I won't be able to see the codes?''}. 
Hence, recording the lectures gives them enough time to catch theory and practice.
It also brings benefits for people who cannot pay attention in class. 
P6 declared, \textit{``I have problems with the speed of things. My brain spins too fast, and things are slow; it makes me anxious. I struggled a lot to pay attention in synchronous classes. Then I watched the recording in 2x, and I could smoothly understand everything''}.




Nevertheless, several participants noted that \textbf{receiving challenging tasks} during the training program was an important aspect of promoting inclusion for developers with disabilities. P11, for example, 
felt included and essential to the team at Catalisa, which did not occur in previous experiences: 
\textit{``I've had previous professional experiences where I felt ableism in everyday life. They only saw me as a quota worker, so they didn't delegate work activities to me. At Catalisa, since the first day, we had challenging goals to achieve in our studies, which made me feel an integral part of the company.''}.

Furthermore, we observed that \textbf{improving participants' work environment} was also perceived as an act of inclusivity. 
Zup provides every new hire with a laptop for work activities, configuring it based on their needs. 
P11 expressed their appreciation for this measure, saying, \textit{``I was amazed when I received the onboarding kit at home, and my laptop was already configured to use NVDA [a screen reader]. I found this reception very wonderful''}. 
In addition, Zup teams' asked participants whether they needed any other adaptations in their work environment. 
P9 mentioned receiving a widescreen monitor, which greatly improved their experience with the computer: 
\textit{``Catalisa, from the beginning, made me feel capable. The company supported me and tried to understand what I needed to participate in the training. They event sent me a large monitor that changed my coding experience''}. 
This improvement was also due to being hired as company employees from the first training day, with a salary and benefits included. As a result, P3 felt \textit{``appreciated, finally having an opportunity in the job market, that improved my family's financial situation''}.

\subsection*{RQ2: \rqtwo}


One of the main challenges faced by participants was the \textbf{lack of accessibility in the programming tools}. 
Particularly for developers with visual impairments, P1 summarized this issue as \textit{``usable but not accessible''} tools, meaning they must constantly switch between different tools to perform simple tasks since the primary tool does not offer them the necessary accessibility. 
For instance, in Insomnia\footnote{https://insomnia.rest/}, they could not edit JSON documents, which is the primary way to manipulate APIs in this tool. 
As a result, participants had a more laborious process of using text editors to edit JSONs and paste the edited version back to Insomnia. 
Two visually impaired people could not use Postman\footnote{https://www.postman.com/} since they could not navigate between the elements in the application using screen readers. 
As P1 remarkably stated, \textit{``This is the life of a blind person; we need to use ten tools to make something.''}.

The participants also faced challenges with IDEs not being accessible to visually impaired people. 
One participant mentioned spending around three months discovering how to change the Java version in IntelliJ. 
The same participant also gave up on using VSCode, since it did not have as shortcuts to use the console, requiring many workarounds to access it. 
The lack of accessibility in IDEs is a well-known issue in the literature. 
Their reliance on visual cues to guide development, such as color highlighting, makes them challenging to use with assistive technologies~\cite{albusays2021diversity}. 
As a result, visually impaired developers often resort to using plain text editors to write code, which can significantly impact activities like editing and debugging~\cite{ehtesham2022grid}. 
Despite the challenges, P1 highlighted the importance of autocomplete features in IDEs for visually impaired developers since they help reduce the cognitive load associated with coding.

Still, we found that tools that provide graphical interfaces for executing \texttt{git} commands are often inaccessible to visually impaired participants, mainly because of the error messages displayed in windows. 
P11 recounted an experience where they lost a coding challenge because they did not receive any warning that their code was not properly committed in \texttt{git}. 
They expected a warning message on the screen, but they are unsure since they could not see it. 
Therefore, despite the recent improvements in programming tools, several participants preferred using the command-line interface, 
as P11 put it, \textit{``
text is the most accessible format''}.

There were also some \textbf{accessibility oversights} during training. 
Participants mentioned teaching habits that can create barriers to inclusiveness. 
For instance, P11 shared that many instructors use IDEs to explain code, frequently using visual instructions such as ``click here'' and ``drag this item here,'' which are inaccessible to individuals with visual impairments: \textit{``As a visually impaired person, I cannot see where to click. I only use the keyboard and shortcuts, and the instructor was not used to this methodology.''}. 
P1 added, \textit{``Starting in programming is already hard, and instructors should not put up additional barriers [due to accessibility oversights].''}. 
Another issue we identified was instructors asking visually impaired individuals to send their code first to \texttt{git}, claiming they could not resolve code merge conflicts. 
P11 deliberately provoked a code merge conflict to learn how to solve it. 
P1 expressed, \textit{``How could they say this for developers with disabilities? 
(...) Also, more complicated issues can be addressed through pair programming. Blind people can code just like everyone else.''}.  



Lastly, we observed that developers with physical impairments \textbf{downplay their efforts} since they may experience fewer challenges related to inclusiveness in remote programming. 
One developer with a physical impairment shared that they often experience back pain due to long work hours, aggravating their spinal condition. Despite this, they considered their disability to be a relatively minor issue when compared to other participants.




\section{Recommendations}

Based on the participants' experience, we propose four evidence-based recommendations for others interested in create Catalisa-like programs.

\vspace{0.2cm}
\noindent
\textbf{Help instructors to master accessibility teaching practices.} Instructors are critical in ensuring that training camps are inclusive for developers with disabilities. 
    They must be familiar with the best accessibility teaching practices and be capable of creating accessible materials such as presentations, codes, and lectures. 
    They must also know how to administer these materials inclusively, e.g., navigating code using assistive technologies such as screen readers. 
    However, many software developers lack expertise in accessibility~\cite{bi2022accessibility}, which makes it challenging to find qualified instructors for these programs. 
    Therefore, it is essential to have companies' support in mastering accessibility teaching practices and listening to participants to understand their needs continually. 
    CS professors already noted this need for accessible materials and how demanding it is to produce them~\cite{huff2021exploring}. 
    The duration of training camps, which can be as short as a few months, only exacerbates the issue since managing inclusive teaching practices requires time and effort. 
    Ultimately, investing in accessibility training for instructors is critical to creating a more inclusive learning environment for developers with disabilities. 

\vspace{0.2cm}
\noindent
\textbf{Make the materials available and organized in advance.} 
    Given the challenges of accommodating everyone's unique needs, many participants suggested that sharing the learning materials in advance can help them better understand the synchronous lectures. 
    For example, visually impaired participants declared that sharing codes snippets with a simple description file aids their comprehension, since they cannot participate in synchronous code reading. 
    Moreover, neurodivergent participants feel less anxious knowing the discussion topics, particularly in slow-paced lectures. 
    They could also consult the material during the class to keep the linearity and maintain others' learning needs. 
    This also allows participants to identify and raise accessibility issues before the synchronous session.

\vspace{0.2cm}
\noindent
\textbf{Curate accessible programming tools.} 
    A selection of programming tools can significantly alleviate the barriers developers with disabilities. 
    However, 
    finding tools that can accommodate everyone's needs can be challenging, which is a well-documented issue in the field of accessibility~\cite{bi2022accessibility,huff2021exploring}. 
    For instance, the participants encountered difficulties with many automated API testing tools they used. 
    While some claimed to be accessible, the participants could not use them smoothly. 
    Finding the right balance between time and resources is crucial to ensure all tools selected during training are accessible.
    By incorporating accessibility considerations from the initial design stages, we can create tools that are inclusive by default and require less retrofitting to meet accessibility requirements.
    
\vspace{0.2cm}
\noindent
\textbf{Make birds of a feather flock together.} 
    Including a person with disabilities in the design and development learning process is crucial for creating a genuinely inclusive environment. 
    In the case of Catalisa, not having someone with a disability on the team meant that they may have missed valuable insights into the challenges and solutions related to accessibility, which could have minimized (or even avoided) some accessibility issues. 
    Moreover, inclusion is not just about providing equal opportunities for people with disabilities but about creating a diverse and inclusive environment with more innovative and effective solutions that work for everyone, regardless of their abilities.


Ultimately, while our recommendations address training camps for developers with disabilities, we argue that an accessible educational setting can benefit everyone. 
For instance, providing materials in advance can help students who may have missed a lecture and facilitates learning~\cite{mackay2019show}, accessibility teaching practices such as describing images can aid those with low internet bandwidth\footnote{https://www.w3.org/standards/webdesign/accessibility}, diverse environments are more understanding~\cite{Vasilescu2015Perceptions}, among other benefits.

\bibliographystyle{IEEEtran}
\bibliography{bibliography.bib}

  

  

  

  

  


\end{document}